\documentclass[namedreferences,hyperref,optionalrh,solaromanenum]{spr-sola}

\usepackage{graphicx}                    
\usepackage{color}                       

\newcommand{\angstrom}{\textup{~\AA} }

\chardef\us=`\_

\graphicspath{{./}{Fig/}}

\begin{document}

\begin{frontmatter}
\title{Coronal Flux Tube Illuminated by Strong Shock Spot: New Year's Eve Solar Eruption of 31 December 2023}

\author[addressref={aff1},corref,email={illya.plotnikov@irap.omp.eu}]{\inits{I.}\fnm{Illya}~\snm{Plotnikov}\orcid{0000-0002-0074-4048}}
\author[addressref={aff1,aff2}]{\inits{A.P.}\fnm{Alexis}~\snm{Rouillard}\orcid{0000-0003-4039-5767}}
\author[addressref=aff3]{\inits{A.}\fnm{Athanasios}~\snm{Kouloumvakos}\orcid{0000-0001-6589-4509}}
\author[addressref=aff4]{\inits{I.~C.}\fnm{Immanuel}~\snm{Jebaraj}\orcid{0000-0002-0606-7172}}

\address[id=aff1]{IRAP, Universit\'e de Toulouse / Observatoire Midi-Pyren\'ees, France}
\address[id=aff2]{Leibniz-Institut für Astrophysik Potsdam, Potsdam, 14482, Germany}
\address[id=aff3]{The Johns Hopkins University Applied Physics Laboratory, Laurel, MD 20723, USA}
\address[id=aff4]{Department of Physics and Astronomy, University of Turku, 20500 Turku, Finland}

\runningauthor{Plotnikov et al.}
\runningtitle{Coronal flux tube illuminated by strong shock spot}

\begin{abstract}
Powerful solar eruptions are known to produce fast and wide shock waves in the solar corona and inner heliosphere. The relationship between the coronal shock waves, solar energetic particles and different types of radio emission is a subject of long-lasting research activity. In this work, we perform a case study of 31 December 2023 eruption that occurred near eastern limb of the Sun. It produced a X5.0 class X-ray flare,  a global EUV wave, a fast $\sim3000$ km/s Coronal Mass Ejection, strong radio emissions (including several type III and type II bursts), solar energetic particles in-situ, and long duration high-energy gamma-ray emission. We employ a technique that combines the reconstructed coronal shock from observations with background coronal MHD simulations to produce shock-mediated synthetic radio spectrum, assuming local emission at plasma frequency. We show that transient high Mach number and quasi-perpendicular coronal shock region explains both a ``hot flux tube'' precursor seen in EUV observations and reverse drifting radio spectral features observed by ground-based facilities. The occurrence of this evanescent strong shock patch was observed when it propagated across pseudo-streamer's cusp where the magnetic field was particularly low. We also find evidence that, at higher coronal altitudes, the low-frequency type II radio burst detected by several spacecraft, is triggered by the interaction of the shock with the heliospheric current sheet. This study provides additional evidence that high-$M_A$ regions of coronal shock surface are instrumental in energetic particle phenomenology.
\end{abstract}
\end{frontmatter}

\section{Introduction}
\label{Sect:Introduction} 

Strong solar eruptions are known for their association with bright flares and for their ability to drive fast and wide collisionless shock waves in the solar corona and beyond. Both flares (along with large and dynamic coronal loops) and shocks are widely recognized as efficient non-thermal particle accelerators. However, the main mechanism and source of the most energetic solar particles which are produced during the early phases of the eruption (when all potential sources are still close to each other in low corona) remain debated. 

Since the advent of multi-point solar observations in remote-sensing imaging \citep[particularly STEREO mission;][]{Kaiser_2008}, it became possible to systematically track and triangulate the propagation of shock waves in the solar corona \citep[e.g.,][]{Kwon_2014, Rouillard_2016, Kouloumvakos_2019}. Radio observations are long known to be valuable diagnostics of shocks in the corona since they can be associated with the electron acceleration at shock waves \citep[see][for recent review]{Klein_2021b}. They are commonly used to track regions of type II emissions and localize where coronal shocks are strong, super-critical and often quasi-perpendicular \citep[e.g., ][]{Zucca_2018, Morosan_2019, Kouloumvakos_2021, Jebaraj_2021}. These studies generally rely on co-temporality and co-location of reconstructed shock wave regions and radio emissions, especially highlighting the region where the shock interacts with a coronal streamer or Heliospheric Current Sheet \citep[e.g.,][]{Cho_2008, Magdalenic_2014, Jebaraj_2020, Frassati_2022, Koval_2023}. 

Radio measurements can also constrain source locations and propagation directions through goniopolarimetric techniques. These methods provide complementary information on the spatial origin and propagation of type II and type III radio bursts  \citep{Fainberg_1972, Reiner_1998, Martinez-Oliveros_2012, Krupar_2012}. Such measurements can help connect radio signatures to the structure of the corona and inner heliosphere.

In a series of recent studies, using spacecraft that had a close magnetic connection to the eruptive region, it was shown that the same part of the shock are responsible for both metric to decametric solar radio emissions and in situ energetic electrons \citep[][]{Morosan_2022, Morosan_2024, Morosan_2025, Jebaraj_2023, Rodriguez-Garcia_2025}. Additionally to evidence of energetic particle acceleration in the complex post-flaring loops beneath the expanding shock \citep[e.g.,][]{Chen_2013, Klein_2005, Klein_2022, Klein_2024}, these studies show that strong shock regions in the low solar corona also accelerate electrons efficiently and are the source of their escape into space as well as precipitation towards the solar surface, depending on magnetic connectivity of the accelerating shock region \citep[see, e.g.,][ for theoretical model]{Mann_2018, Wu_2021}.

In this work, we study the solar eruption of 31 December 2023. Several recent works addressed different aspects of this event: localization of hard X-ray emissions \citep{Ryan_2024}, imaging of the plasma blob ejection \citep{Roy_2025}, coronal wave interaction with coronal hole \citep{Zhang_2025}, properties of herringbone radio emissions \citep{Zhang_2026}. The particularity of the present work is that we focus on the coronal shock wave evolution by combining Extreme Ultraviolet (EUV) and coronagraphic imaging, addressing the relation between the shock properties and observed radio emissions.” For this event, the radio observations showed that the type II radio burst exhibited an unusual hotspot with a reverse drift and EUV images showed a transient precursor brightening. Both features are not typical and we further explore possible explanations.

Using multi-point EUV and white-light observations we reconstruct the spatial evolution of the coronal shock, which is found to be very fast; reaching 3500 km/s at peak. The reconstruction is then coupled with MHD simulation of the solar corona. 
Using the technique of \citet{Kouloumvakos_2021} and \citet{Jebaraj_2021} that produces shock-mediated synthetic radio spectrum, we present evidence that one evanescent particularly strong shock ``patch'' along its Earth-directed flank was responsible for ``hotspot'' radio emission feature and for the transient brightening of a localized upstream flux tube observed in EUV.

The article is organized as follows. In \autoref{Sect:Obs} we present an overview of the solar eruption observations. \autoref{Sect:3D_Reconstruction} presents the coronal shock reconstruction using remote-sensing EUV and white-light observations. In \autoref{Sect:Results} we present the results combining observations and shock modeling. Discussion and conclusions are presented in \autoref{Sect:Conclusion}.   

\section{Overview of the Solar Eruption}
\label{Sect:Obs} 

\begin{figure}    
\centerline{\includegraphics[width=0.95\textwidth]{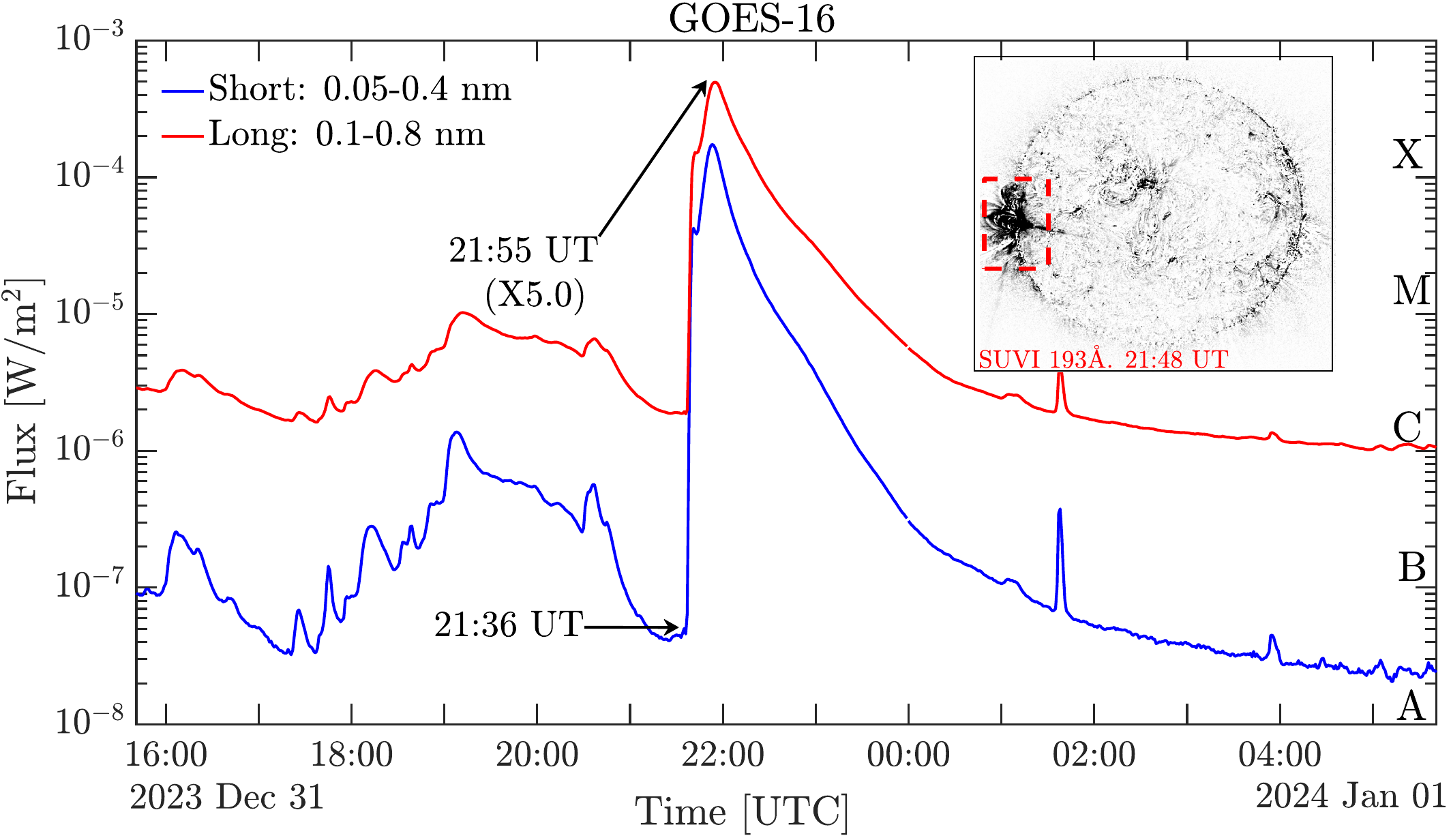}}
\small
  \caption{GOES-16 soft X-ray observations from 2023/12/31 to 2024/01/01. The X5.0 flare started at 21:34 UT and peaked at 21:55 UT (indicated by arrows). The inset image displays the running-difference image at 193\angstrom wavelength from 21:48 UT and 21:44 UT images, taken by Solar Ultraviolet Imager \citep[SUVI;][]{Darnel_2022} onboard GOES-16. The flaring region at the eastern edge of the Sun is highlighted with red dashed box.
                }
\label{fig:GOES16_lightcurve}
\end{figure}

On 31 December 2023, a major solar eruption developed in association with NOAA active region (AR) 13536, located at Stonyhurst heliographic latitude $+4^\circ$ and longitude $-76^\circ$ (N04E76) as seen from Earth viewpoint. The associated soft X-ray flare was observed by Geostationary Operational Environmental Satellite 16 \citep[GOES-16;][]{Garcia_1994}. It was of X5.0 class, beginning at approximately 21:36 UT and peaking at 21:55 UT (see the light-curve in  \autoref{fig:GOES16_lightcurve}).

\begin{figure}    
\centerline{\includegraphics[width=1.0\textwidth]{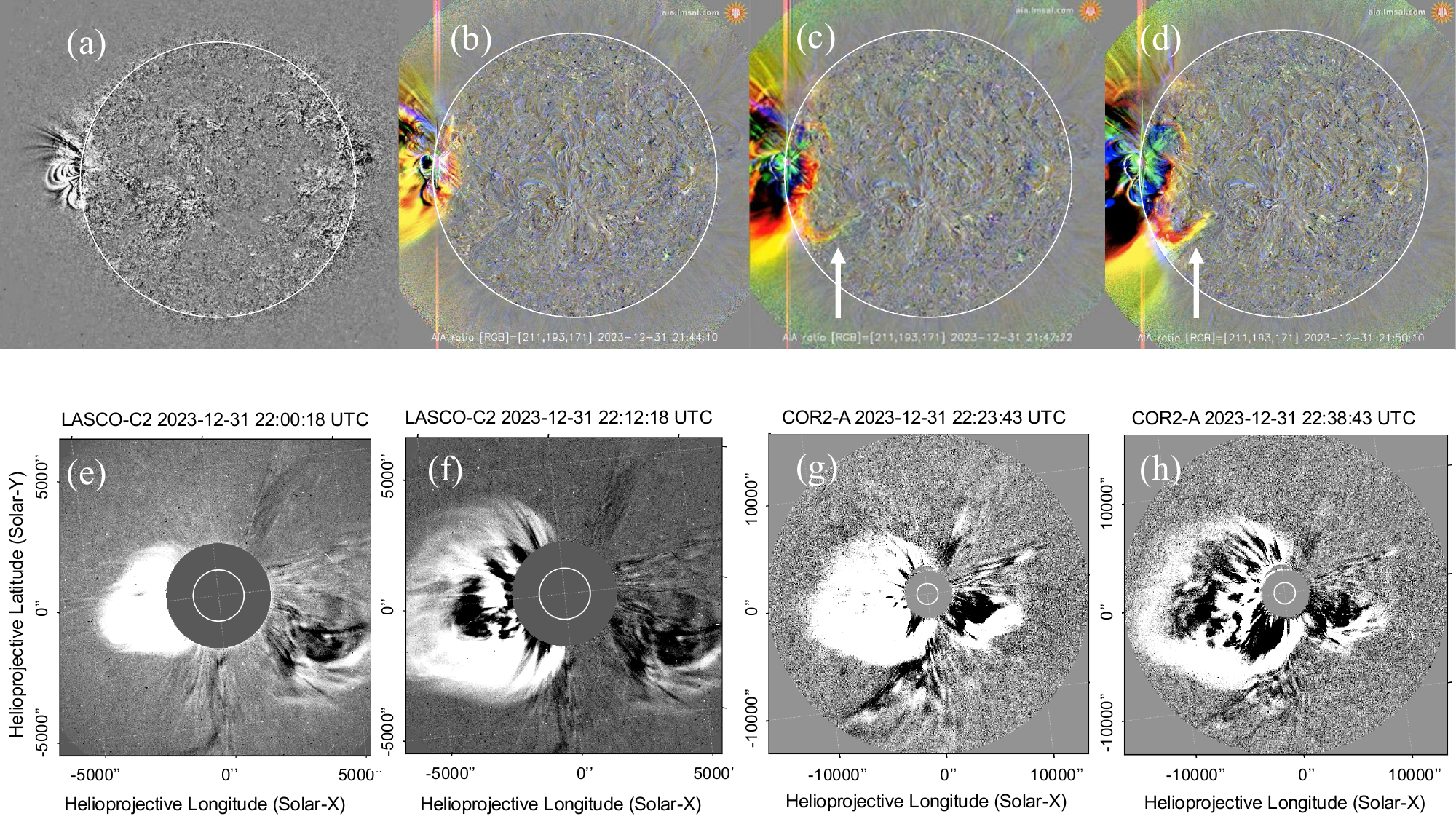}}
\small
  \caption{Selection of running-difference images from STA/EUVI at 195\angstrom (a), SDO/AIA at [211,193,171]\angstrom (b-d), LASCO C2 (e,f), and STA/COR2 (g-h) at different times: 21:42 UT, 21:44 UT; 21:47 UT, 21:50 UT, 22:00 UT, 22:12 UT, 22:24 UT and 22:39 UT. The edge of the solar disk is delimited by white circle in each panel.}
\label{fig:EUV_C2_obs}
\end{figure}

As common for largest flares, it was accompanied by a global coronal pressure wave surrounding a set of bright coronal loops \citep{Veronig_2010, Warmuth_2015}. The wave expansion was observed in EUV by Atmospheric Imaging Assembly \citep[AIA;][]{Lemen_2012} on board of Solar Dynamics Observatory \citep[SDO;][]{Pesnell_2012} and by Extreme Ultraviolet Imager \citep[EUVI;][]{Howard_2008} on board of Solar TErrestrial RElations Observatory Ahead \citep[STEREO-A or STA;][]{Kaiser_2008} mission. Briefly after the initial phase of the flare, a fast Coronal Mass Ejection (CME) lift-off was seen by C2 and C3 coronagraphs of the Large Angle Spectroscopic Coronagraph \citep[i.e., LASCO-C2 and C3;][]{Brueckner_1995} on board Solar and Heliospheric Observatory (SOHO) mission, and the COR2 coronagraph of the STA mission \citep[i.e., STA/COR2;][]{Howard_2008}. The CME first appeared in LASCO/C2 between 21:48 UT and 22:00 UT and in STA/COR2 field-of-view between 21:38 UT and 21:53 UT. A selection of STA/EUVI (panel \emph{a}) AIA (panels \emph{b-d}), LASCO-C2, (panels \emph{d-e}), and STA/COR2 (panels \emph{f-g}) observations is presented in \autoref{fig:EUV_C2_obs}. The leading edge of the CME crossed the entire field of view of STA/COR2 in less than one hour, making the projected velocity larger than 2400 km/s. According to the SOHO/LASCO CME online catalog\footnote{\url{https://cdaw.gsfc.nasa.gov/CME_list/}} \citep{Yashiro_2004, Gopalswamy_2024} the plane-of-sky linear speed of the CME was 2852 km/s. According to the full LASCO catalog, it was among five of the fastest CMEs ever recorded by the instrument since 1996 and the fastest of Solar Cycle 25 (at least until April 2025)\footnote{Of course, this does not take into account the events that were hidden to LASCO and/or propagating closely to Sun-Earth direction.}.
One particular feature that we will focus on is visible in panels \emph{c,d} of \autoref{fig:EUV_C2_obs}: a bright loop ahead of the expanding wave, located by a white arrow. It is seen to connect upstream of south-western flank of the expanding wave closer to the center of the solar disk between 21:47 UT and 21:53 UT in SDO/AIA and STA/EUVI images, before being fully engulfed by the expanding coronal wave.

\begin{figure}    
\centerline{\includegraphics[width=1.0\textwidth]{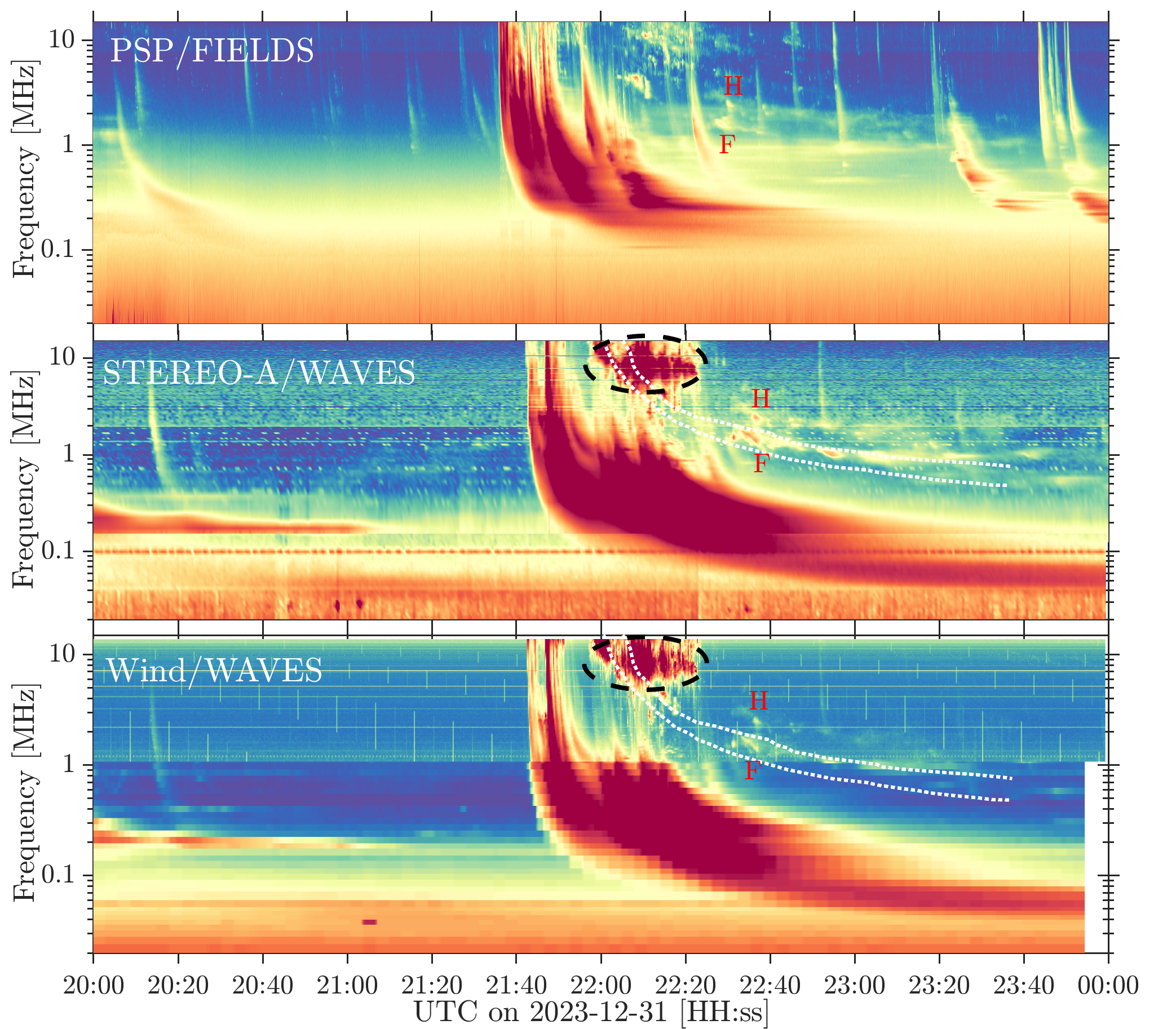}}
\small
  \caption{Dynamic radio spectrum observed by PSP/FIELDS (top panel), STEREO-A/WAVES (middle panel) and Wind/WAVES (bottom panel) instruments from 20:00 UT to 00:00 UT on 31 December 2024, in spacecraft time (no shift applied for Light Travel Time differences). The labels `F' and `H' refer to what appear to be fundamental and harmonic emission, respectively. Two white dotted lines are from the shock reconstruction and modeling presented in \autoref{fig:synthetic_spectrogram} and described in detail in  \autoref{Sect:Results}.
                }
\label{fig:Radio_PSP_STA_Wind}
\end{figure}

The eruption was also rich in radio emissions. In \autoref{fig:Radio_PSP_STA_Wind} we present dynamic radio spectra observed by FIELDS instrument \citep{Bale_2016} on board of Parker Solar Probe (PSP), S/WAVES instrument \citep{Bougeret_2008} on board of STA, and WAVES instrument \citep{Bougeret_1995} on board of Wind mission from top to bottom panels, respectively. The spectral range of each panel in \autoref{fig:Radio_PSP_STA_Wind} is between 20 kHz and 15 MHz (common frequency range across all three instruments) and the time window four hours from 20:00 UT on 2023-12-31 to 00:00 UT on 2024-01-01. The event started at 21:36 UT with a type III burst recorded by PSP/FIELDS. First type III burst at STA and Wind was recorded several minutes later, at 21:42 UT at the highest instrument frequency of 15 MHz. All type III bursts show characteristic fast drifting pattern from the upper instrumental frequency (15 MHz) to below 100 kHz. Several other type IIIs occurred at each observatory up to 22:15 UT, indicating episodes of accelerated electron beams escaping into the the interplanetary medium. The sequence of bursts was clearly different at PSP as compared to Wind and STA. We note that it was an east-side limb flare as seen from Earth and STA and a slightly behind the western limb as seen from PSP (located at 0.17 AU from the Sun and separated from Earth by about 167° in heliographic longitude, i.e. nearly in superior conjunction). It was therefore observing the opposite flank of the eruption. For the very first burst, if we assume that the $\sim 15$ MHz emissions occurred near the flaring region, the delay of 7 min can be accounted for by the source-spacecraft light-travel time \footnote{This estimate neglects the possibility of different source regions of the burst seen by PSP and near-Earth spacecrafts, as well as the propagation effects of emitting electron beams.}. 

At least one type II burst was observed by each observatory. It started at nearly 21:58 UT at 15 MHz at STA and Wind, and slightly later at PSP. Slowly drifting emission is seen until the end of displayed observations at 00:00 UT on 2024-01-01. It has characteristic patchy features and well-identifiable second harmonic emission. We performed a linear fit using the frequency drift rate and assuming either Mann \citep[][]{Mann_2023} or Leblanc \citep[][]{Leblanc_1998} coronal density model. According to the fit, the linear radial propagation velocity of emitter was $2000 \pm 300$ km/s (Mann model) or $1200 \pm 200$ km/s (Leblanc model) at Wind and STA. Slightly slower velocity was obtained for type II event at PSP/FIELDS.

A patch of intense emissions was observed by STA/WAVES and Wind/WAVES from 22:00 UT to 22:20 UT in frequency band between 5 MHz and 15 MHz. The corresponding zone is highlighted in middle and lower panels of \autoref{fig:Radio_PSP_STA_Wind} using black dashed circles. Their relationship to the principal type II feature is not unambiguous from the dynamic spectrum alone. We therefore describe them as additional fine structures, resembling herringbones \citep{Cane_White_1989}, rather than assigning them definitively to a separate emission component. Their interpretation would require further analysis of their drift rates, morphology, and temporal association with the principal type II burst. The fine-structure of individual pulses can be better captured by large ground based observatories when involved emission occurs at slightly higher frequencies \citep[as shown in][for recent studies]{Morosan_2019, Morosan_2024, Magdalenic_2020, Koval_2023, Jebaraj_2023, Zhang_2026}.

\begin{figure}    
\centerline{\includegraphics[width=0.99\textwidth]{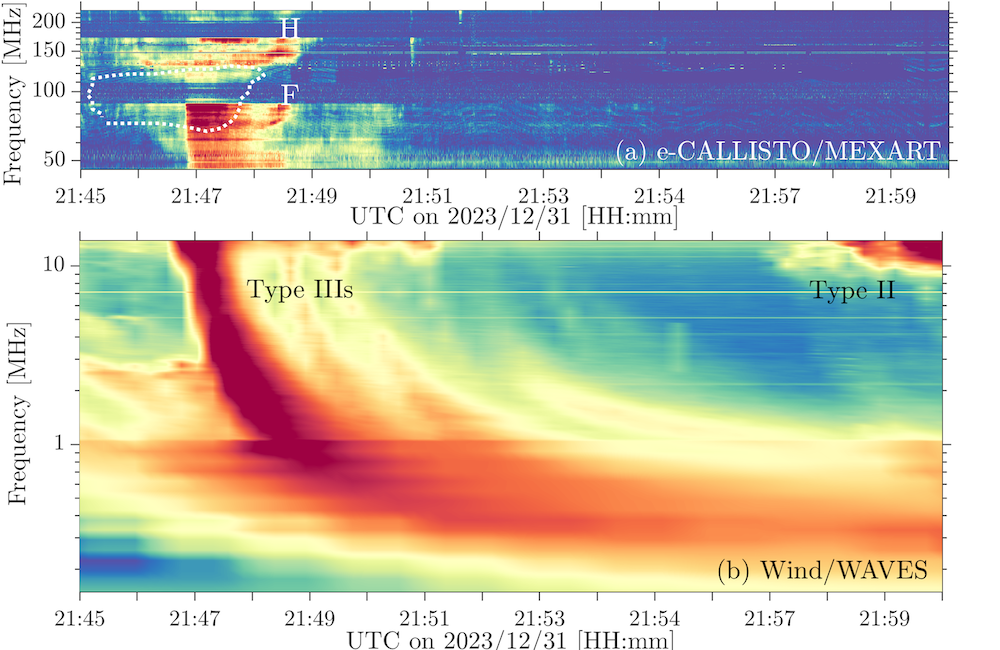}}
\small
  \caption{Composite dynamic radio spectrum on 31 December 2023 at frequencies 0.15 - 230 MHz. It consists of spectra from e-CALLISTO MEXART (upper panel) and Wind/WAVES (bottom), from 21:45 UT to 22:00 UT. The labels `F' and `H' refer to what appear to be fundamental and harmonic emission, respectively. The region delimited by white dotted line is the ``hotspot'' from the shock reconstruction and modeling, described in detail in \autoref{Sect:Results}.
                }
\label{fig:Radio_eCallisto}
\end{figure}

Additionally to spacecraft measurements, an interesting radio burst was observed by several ground-based e-CALLISTO radio observatories \citep[][]{Benz_2009, Monstein_Csillaghy_Benz_2023}\footnote{\url{http://www.e-callisto.org/}} at $>40$ MHz frequencies, including Alaska-HAARP, Australia-ASSA, MEXART. In \autoref{fig:Radio_eCallisto} we present a composite radio spectrum from e-CALLISTO/MEXART (top panel) at 40-230 MHz frequency range and from Wind/WAVES (bottom panel). The time window is smaller than \autoref{fig:Radio_PSP_STA_Wind} ranging from 21:45 UT to 22:00 UT, allowing to better resolve the high-frequency feature observed by MEXART between 21:47 UT and 21:50 UT. This feature consists of an initial patch between 21:47 UT and 21:48 UT covering all frequencies between 40 MHz to 80 MHz. Then a narrow-band reverse (i.e. towards higher frequencies) drifting feature is observed from 70 MHz at 21:48 UT to 85 MHz at 21:49 UT. A harmonic emission is also clearly drifting from $\approx130$ MHz to $\approx160$ MHz. The emission weakens after 21:49 UT and disappears after 21:50 UT. We also attempted to fit the radio feature using the Mann and Newkirk coronal density models. The fit provided a radially downward-propagating emitter with $900 \pm 200$ km/s for both models.

The metric feature observed by e-CALLISTO between approximately 21:47 and 21:50 UT overlaps in time with the onset of the type III emission detected by Wind/WAVES. This temporal coincidence does not, by itself, establish whether the features originate from the same electron population or from distinct emitting regions. The association of the metric lanes with fundamental and harmonic emission is therefore assessed using their frequency–time morphology and the relationship between the lanes, rather than timing alone. An alternative interpretation is that some of the observed structure belongs to the type II emission detected at lower frequencies by Wind/WAVES at later times as shown in bottom panel of \autoref{fig:Radio_eCallisto}. We discuss this ambiguity when comparing the observations with the synthetic radio spectrum in \autoref{Sect:Results}.


\emph{High-energy gamma-ray emissions:} Fermi/LAT detected solar-origin gamma-ray emission above 100 MeV from approximately 22:10 UT on 31 December 2023 until 00:23 UT on 1 January 2024 \footnote{According to Fermi Solar Flare Observations facility available at \url{https://hesperia.gsfc.nasa.gov/fermi_solar/}}. Being of nearly two hour duration puts this event at the limit of late-phase (or long-duration) gamma-ray event, as defined by \citet{Share_2018}. The duration of this emission is also comparable to that of the interplanetary type II radio burst. We note that the temporal correlation is consistent with a possible relationship between the evolving CME-driven shock and the sustained gamma-ray event \citep[e.g., ][]{Gopalswamy_2018}, but it does not establish a common origin \citep[see discussion in][Section 4]{Klein_2024}. In particular, the present radio and shock reconstruction does not determine whether the gamma-ray-producing protons were accelerated by the shock (or underlying structures closer to the erupting active region), how they were transported, or where they interacted to produce gamma rays. A quantitative assessment of the required proton population and its reservoir would require dedicated gamma-ray spectral modeling, as well as particle acceleration and transport calculations, which are beyond the scope of this study.

\emph{Solar energetic particle measurements in-situ:} A solar energetic particle (SEP) event was observed by instruments onboard spacecrafts that were largely separated in longitudes (L1, STA, Solar Orbiter, Bepi Colombo, PSP, Mars). The SEP event and onset linked to our eruption was very clear at PSP, Bepi Colombo\footnote{Marco Pinto, private communication} and in martian environment as measured by MAVEN and TGO-Liulin \citep{Semkova_2025}. Magnetic connection was poor for L1, STA and Solar Orbiter in addition to high pre-event fluxes at their locations because of SEP event originating from earlier eruption on the western side of the Sun. A dedicated study relating the shock and SEP properties from this event will be done in a separate work.

\section{Shock reconstruction}
\label{Sect:3D_Reconstruction} 

In order to reconstruct the three-dimensional (3D) kinematic properties of the coronal shock we follow the method developed by \citet{Rouillard_2016} and improved in a series of following studies \citep[e.g.,][]{Kouloumvakos_2019, Kouloumvakos_2022b, Jarry_2024}. It relies on the ellipsoid model of the expanding coronal pressure wave \citep{Kwon_2014} and on established techniques of shock identification in EUV and white-light observations \citep[e.g.,][]{Vourlidas_2003, Vourlidas_2009, Rouillard_2012}. When available, simultaneous well-separated viewpoint observations are used to constrain the 3D extension of the shock at different times. For this event, we used observations from L1 and STEREO-A viewpoints during early expansion (before 22:15 UT), using data from SDO/AIA, SOHO/LASCO coronagraphs, STA/EUVI and STA/COR2.  \\

This type of analysis was made more challenging by the fact that Earth and STA were separated by 7 degrees in longitude. Both had lateral view of the CME expansion from nearly the same perspective. Later expansion (after 22:20 UT) was also observed by PSP/WISPR \citep[][]{Vourlidas_2016} when the apex of the structure was far enough to be seen in Heliospheric Imagers. By the time of the event, PSP was located nearly on the opposite side of the Sun (193 degrees longitudinal separation with Earth) at radial distance of $0.17$AU or $36.5 R_\odot$. It provided a valuable viewpoint to constrain the global shape of the evolved structure, as it had nearly a lateral view from the opposite side. For its later evolution (after 23:30 UT; not reported here) we also used STA/HI1 and SOLO/HI observations to check that the self-similar evolution from earlier fittings was consistent with late-time expansion into the inner heliosphere. The analysis was performed using \emph{PyThea} shock wave analysis software\footnote{\url{www.pythea.org}} \citep[][]{Kouloumvakos_2022b} from 21:42 UT to 23:30 UT.

\begin{figure}    
\centerline{\includegraphics[width=0.99\textwidth]{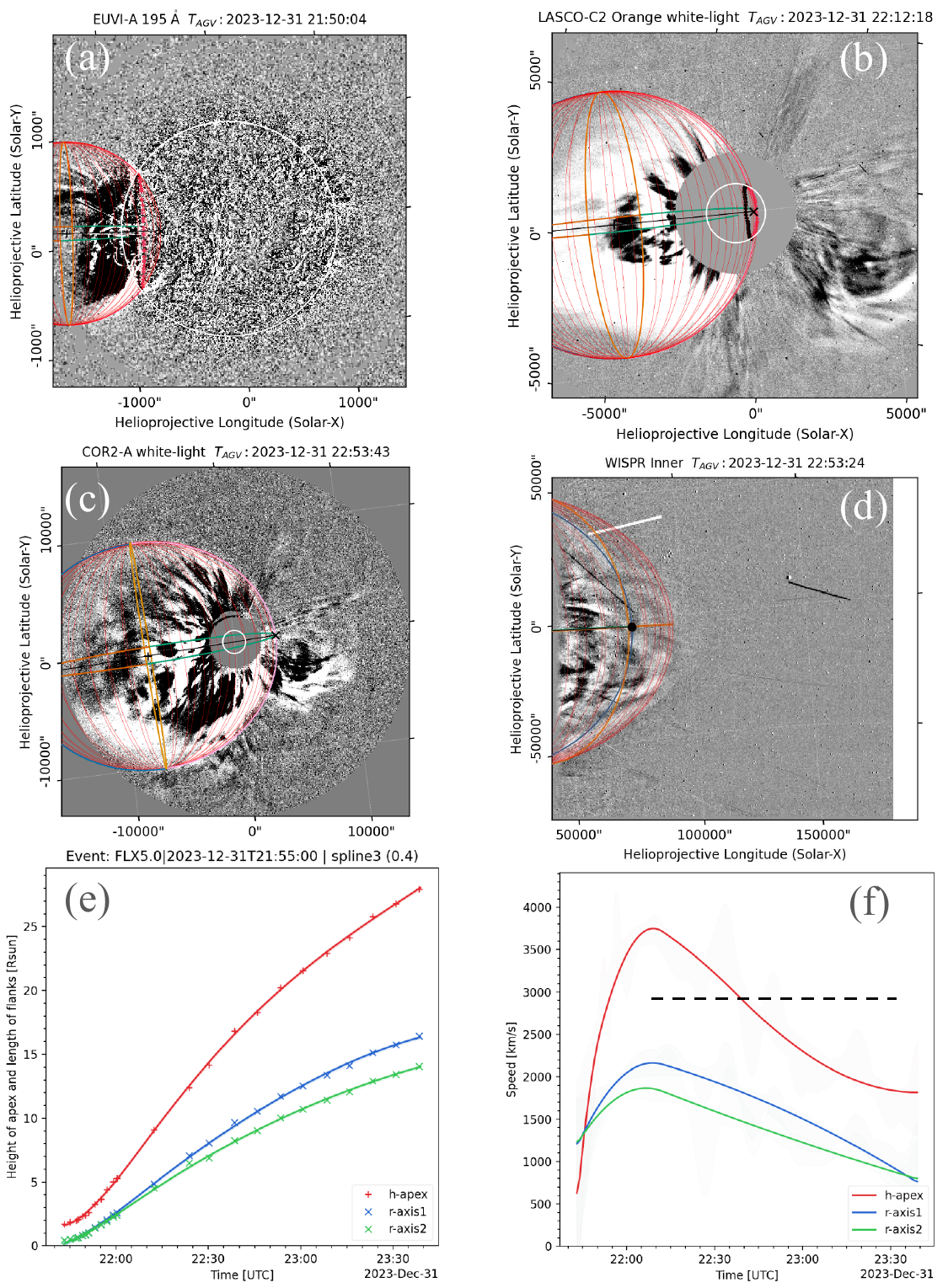}}
\small
  \caption{Results of the coronal shock reconstruction. Panel a: STA/EUV running-difference image at 21:50 UT with the fitted shock surface shown using red-colored mesh. Panel b: same but for LASCO-C2 image at 22:12 UT. Panels c-d: reconstructed shock on STA/COR2 and PSP/WISPR-IN images at 22:53 UT. Panel c: heights of the apex and ortho-radial axes of the fitted shock at all times used for reconstruction. Panel f: derived speeds. Dashed horizontal line shows the linear speed value from SOHO/LASCO CME catalogue discussed in \autoref{Sect:Obs}.
                }
\label{fig:shock_reconstruction}
\end{figure}

In \autoref{fig:shock_reconstruction}, we present an overview of the shock reconstruction results. Panels \emph{a-d} show the running-difference images of STA/EUVI at 21:45 UT, LASCO/C2 at 22:15 UT, STA/COR2 and PSP/WISPR at 22:53 UT, respectively. The best fit of the shock is over-plotted using a red mesh. In panel \emph{e}, we show the time-evolution of heights of the structure at the sequence of reconstructed times and in panel \emph{f} the derived time-evolution of the corresponding velocities.
During the early evolution, we assumed that the shape of the structure was close to a spheroid, that is statistically a reasonable approximation \citep{Jarry_2023}. Thus, constraining the lateral expansion of the structure from observations provided a reasonable estimate of the ellipsoid size in the line-of-sight direction.  Later on, when PSP/WISPR observation could be used, the structure appeared to gradually evolve into slightly oblate ellipsoid (see difference between blue and green lines in panels \emph{c} and \emph{d}).

It appears that the shock had rapid expansion into the solar corona exceeding shortly $3500$ km/s around 22:10 UT and then decelerating. The average  transit speed through the corona is consistent with the value given by the LASCO catalogue, indicated in \autoref{fig:shock_reconstruction}(d) by the horizontal dashed line (2852 km/s, see quote in \autoref{Sect:Obs}). The value from LASCO/CME catalogue corresponds to a linear fit when the shock height from the solar photosphere was between 5 $R_{\odot}$ (at 22:00 UT) and 25 $R_{\odot}$ (at 23:18 UT). 

In the following step, we combined the expansion of the shock with a background MHD model of the corona. As in previous studies \citep[][]{Rouillard_2016, Plotnikov_2017, Kouloumvakos_2019, Kouloumvakos_2022a, Jarry_2024}, we used publicly available simulation provided by Predictive Science Inc.  \citep{Lionello_2009, Riley_2011} employing the Magnetohydrodynamic Algorithm outside A Sphere Thermodynamic (MAST) code. It is a 3D MHD model that uses SDO/HMI magnetograms as inner boundary input of the magnetic field and employs advanced thermodynamics for coronal conduction and parametrized coronal heating. It was proven to accurately reproduce coronal plasma and magnetic field conditions in the corona \citep{Lionello_2009, Riley_2011}\footnote{\url{https://www.predsci.com/mhdweb/data_access.php}}. The result of this combination enables the derivation of various parameters of the shock, such as Mach numbers, compression ratio, shock-normal obliquity with respect to the upstream magnetic field direction as well as tracing magnetic connectivity inside solar corona at radial distances extending from solar surface to $30 R_\odot$. We did not apply any rescaling factors to magnetic field intensity or density provided by the simulation output. The influence of this choice will be discussed in \autoref{Sect:Conclusion}.

\section{Results}
\label{Sect:Results} 

\begin{figure}    
\centerline{\includegraphics[width=0.99\textwidth]{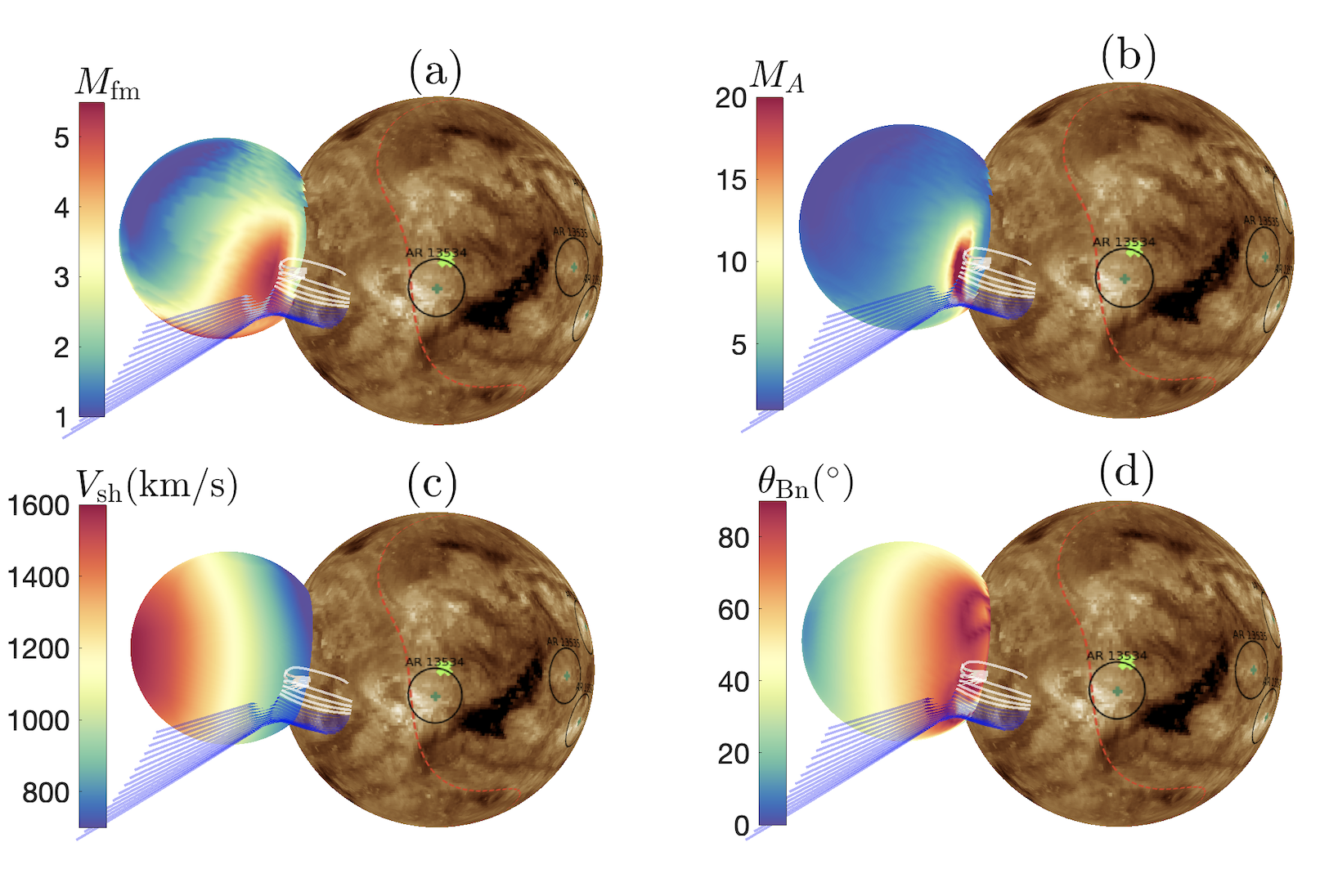}}
\small
  \caption{Fitted shock surface on 2023-12-31 at 21:47 UT as seen from Earth position. The color-coding shows the distribution of fast-magneto-sonic Mach number $M_{\rm fm}$ (panel a), Alfv\'en Mach number $M_A$ (panel b), shock front speed (panel c), and the angle between shock-normal and the upstream magnetic field vector $\theta_{\rm Bn}$ (panel d).  Solar surface is mapped with 193\angstrom EUV image recorded by SDO/AIA. A selection of field lines crossing the high $M_{\rm fm}$ and $M_A$ regions is plotted using blue and white lines corresponding to open and closed field lines, respectively. Red dotted line across solar surface delimits the magnetic field inversion line (HCS). Sub-Earth radial projection point is indicated by a green 'x' symbol and NOAA active regions are located by black circles with dark green cross in the center.
                }
\label{fig:hotspot_EUV_shock_highMA}
\end{figure}

In \autoref{fig:hotspot_EUV_shock_highMA}, we present the distribution of different quantities on the reconstructed shock surface at 21:47 UT as seen from Earth point-of-view. Panels \emph{a-d} show the fast magneto sonic Mach number $M_{\rm fm}$, Alfv\'en Mach number $M_A$, shock front speed $V_{\rm sh}$, and shock obliquity angle $\theta_{\rm Bn}$ distributions, respectively. The solar surface is mapped with EUV image from SDO/AIA at 193\angstrom wavelength. \\

For this early time during the eruption, the shock apex was already fast (exceeding 1600 km/s), but most of the surface of the pressure wave was not yet forming a strong shock ($M_{\rm fm} <2$), except a clearly visible region at the south-western flank where $M_{\rm fm}>6$, $M_A>20$, $V_{\rm sh} \approx 950$ km/s, and $\theta_{\rm Bn} \approx 80^\circ - 90^\circ$. More detailed inspection of the magnetic configuration of this region showed that it corresponds to the cusp of a pseudo-streamer region where the magnetic field strength drops significantly, favoring locally the formation of a strong shock (speed exceeding local $V_A \simeq 100$ km/s). This shock region was quasi-perpendicular, favoring rapid electron acceleration by SDA mechanism \citep{Mann_2018, Wu_2021, Amano_2022}. Magnetic connectivity lines from this ``hotspot'' are shown by blue (open) and white (closed) lines in \autoref{fig:hotspot_EUV_shock_highMA}.

\begin{figure}    
\centerline{\includegraphics[width=0.84\textwidth]{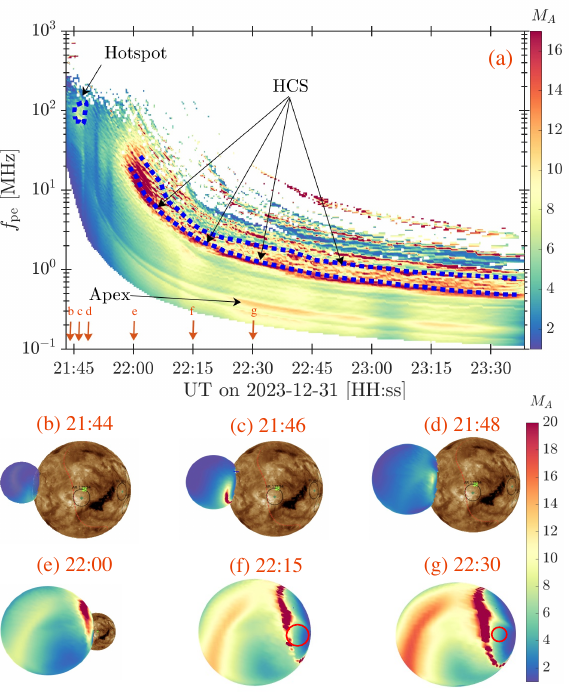}}
\small
  \caption{Synthetic radio spectrum based on shock $M_A$ values (panel a). Bottom panels (b-g): snapshots of $M_A$ distribution on fitted shock surface, as seen from Earth viewpoint, at different selected times (indicated by red arrows in upper panel). In two last sub-panels at 22:15 UT (f) and 22:30 UT (g) the position of solar disc (hidden under the expanding shock) is shown by a red circle. Two dotted blue lines follow the evolution of low- and intermediate-frequencies regions of shock-HCS interaction (dark red zones corresponding to $M_A >14$). These two lines are reported in \autoref{fig:Radio_PSP_STA_Wind} (using white color) on top of STA/Waves and Wind/WAVES spectra. The region around the bright region at the upper left corner, labeled as ``hotspot'', is encircled using blue dotted line. It is reported into \autoref{fig:Radio_eCallisto}, upper panel. 
                }
\label{fig:synthetic_spectrogram}
\end{figure}

According to \citet{Kouloumvakos_2021} the Alfv\'en Mach number is a convenient indicator of which regions of the shock surface are most efficient at producing type II radio emissions in conjunction with a favorable shock geometry from oblique to quasi-perpendicular. Our procedure follows the same approach as in \citet{Kouloumvakos_2021} and \citet{Jebaraj_2021}. This method projects each part of the shock surface onto frequency-$M_A$ space and then selects the mean value of $M_A$ for each frequency\footnote{In the original method of \citet{Kouloumvakos_2021} they selected the median value.}. Performing this at multiple time-steps provides the shock-mediated dynamic spectrum. We note that synthetic radio frequencies are assigned using the local electron density from the MHD MAST simulation along the reconstructed shock surface, assuming plasma emission at the local electron plasma frequency ($f_{\rm pe} = 8.98 \sqrt{N_e /  1 {\rm cm}^{-3}}$ kHz). These frequencies therefore depend directly on the accuracy of the modeled density distribution. No global density rescaling to MAST simulation was applied in the reference calculation, which seems a reasonable choice after comparing the radial profiles extracted from MAST simulation with available in-situ data at 1 AU and at 30 $R_{\odot}$ (PSP measurements). The implication of this choice is discussed later in article, in \autoref{Sect:Conclusion}. 

In \autoref{fig:synthetic_spectrogram} we show the synthetic radio spectrum (upper part, panel a) that maps $M_A$ value of different regions of the shock into time-frequency map according to the procedure described above. The lower part (panels \emph{b-g}) shows the distribution of $M_A$ on the shock surface at six selected times: 21:44, 21:46, 21:48, 22:00, 22:15, and 22:30 UT. The solar surface is presented with 193\angstrom map in each panel (as in \autoref{fig:hotspot_EUV_shock_highMA}). It appears that the hotspot of nearly 3 min duration (21:45 UT to 21:48 UT) was located on the lower Western part of the expanding shock. Later evolution shows that high-$M_A$ regions correspond to the interaction of the shock with the HCS. The position of the HCS can be readily intuited by inspecting the orientation of the red dashed line on solar surface in \autoref{fig:hotspot_EUV_shock_highMA} (magnetic inversion line that evolves at large radii into coronal streamer and the HCS). Hence, different dark red lines in panel a correspond to different emission regions of the shock interaction with the HCS. We also notice that the region around the shock nose (i.e., labeled as `Apex' in the figure) appears as weaker emitter. Despite its high propagation speed, the shock apex is predicted to contribute relatively weakly to the synthetic type II emission. In the model, this region has a lower Alfv\'en Mach number and a more quasi-parallel shock geometry than the flanks interacting with the HCS. These properties may reduce the efficiency of electron acceleration and help explain why the apex does not produce a comparably prominent type II signature in the observed dynamic spectra. This interpretation concerns the type II emission and should not be taken as an explanation for the separate type III bursts.
 
The synthetic spectrum can be used to directly compare our modeling with observations. Two white dotted lines in \autoref{fig:synthetic_spectrogram}a) follow two defined high-$M_A$ regions that correspond to regions of shock interaction with the HCS. These two lines are reported in \autoref{fig:Radio_PSP_STA_Wind}(b,c) where STA/Waves and Wind/WAVES radio observations are shown. It appears that the observed type II burst signatures are located in between these two lines, slightly closer to the upper (intermediate frequency shock-HCS interaction line). The hotspot related region is encircled by white-dotted line in upper left corner and is reported in \autoref{fig:Radio_eCallisto}(a) that shows MEXART observations. We note that the modeled region starts roughly 1 min earlier than the bulk of observed radio emission. However, the bandwidth and duration of the emission is reasonably consistent with modeled hotspot region.  

\subsection{Input from fast shock acceleration theory of electrons}
\label{subsect:SDA_synthetic_spectra}

The Mach number of the shock is not the only relevant quantity that tracks radio-producing regions at the shock. According to established theory of type II radio bursts, the emission occurs from interaction of shock-accelerated electron beams and local upstream plasma \citep[e.g.][]{Mann_2005}. The shock has to be locally highly oblique and fast in order to form fast enough electron beams near the front, driving local Langmuir wave growth. According to \citet{Mann_2018}, the pre-condition triggering the instability reads as $V_b / (\sqrt{3}V_{th,e}) >1$. Where $V_b = V_{\rm sh}(1+\cos \alpha_{\rm lc}) /\cos \theta_{\rm Bn}$ is the speed of the beam,  $V_{th,e}$ is the local coronal electron thermal speed, and the loss-cone angle is related to the overshoot value of magnetic field at the shock front as $\alpha_{\rm lc} = \arcsin{\sqrt{B_1 / B_{\rm ov}}}$. The produced turbulence of Langmuir waves is then partly converted into electromagnetic emission at local plasma frequency and/or its harmonics. Adopting the theoretical formalism of \citet{Mann_2018, Mann_2022}, we diagnose which regions of the expanding coronal shock are able to trigger the instability with high enough number density of accelerated electrons to power radio emissions. For this purpose, we used three derived quantities considered as representatives of shock-acceleration efficiency of electrons and emission intensity in radio waves:

\begin{enumerate}
\item Instability criterion: $L_c = V_b / (\sqrt{3}V_{th,e}) $. As mentioned earlier, this quantity has to be larger than $1$ in order to produce the beam-driven instability driving Langmuir waves. Mapping of $L_c$ on the whole surface of the shock was firstly done by \citet{Kouloumvakos_2021} in the study of 05 November 2014 eruption.
\item Fraction of electrons that is accelerated by reflection on the shock front $N_{\rm acc} / N_1$, derived according to Equation (32) in \citet{Mann_2022}. The shock reconstruction coupled with MHD coronal model provides all necessary information for this calculation, assuming upstream electron temperature is equal to proton temperature, $T_e = T_p = T_1$.
\item Efficiency of conversion of beam kinetic energy into radio waves, defined as $\varepsilon_{\rm rad} = Q W_b / W_{\rm radio}$. This parameter combines the previous two, as the kinetic energy density of the beam is $W_b = 0.5 N_{\rm acc}  m_e V_b^2$. Here, we adopted the conversion factor $Q = 10^{-6}$ which is a conservative value. It is three times smaller than derived in original study of \citet{Ginzburg_1958} as recent fully-kinetic simulations indicate that only a small fraction of produced transverse waves is escaping as O-mode and X-mode emission \citep{Krafft_2025}, while larger fraction is contained near the source in form of Z-mode waves. Yet, conversion efficiencies are always higher than $Q =10^{-6}$ \citep[see Figure 2 in][]{Krafft_2025}. Following \citet{Mann_2022} (Equation 34), when estimating $W_{\rm radio}$ we assumed that the type II burst observed at 1 AU has a typical flux of $10^3$ sfu and then rescaled the radio emissivity down to the Sun (or emission distance of the shock when it is far from the Sun). The bandwidth of $\Delta f = 10$ MHz was assumed, typical for type II bursts.
\end{enumerate}

\begin{figure}
\centerline{\includegraphics[width=0.99\textwidth]{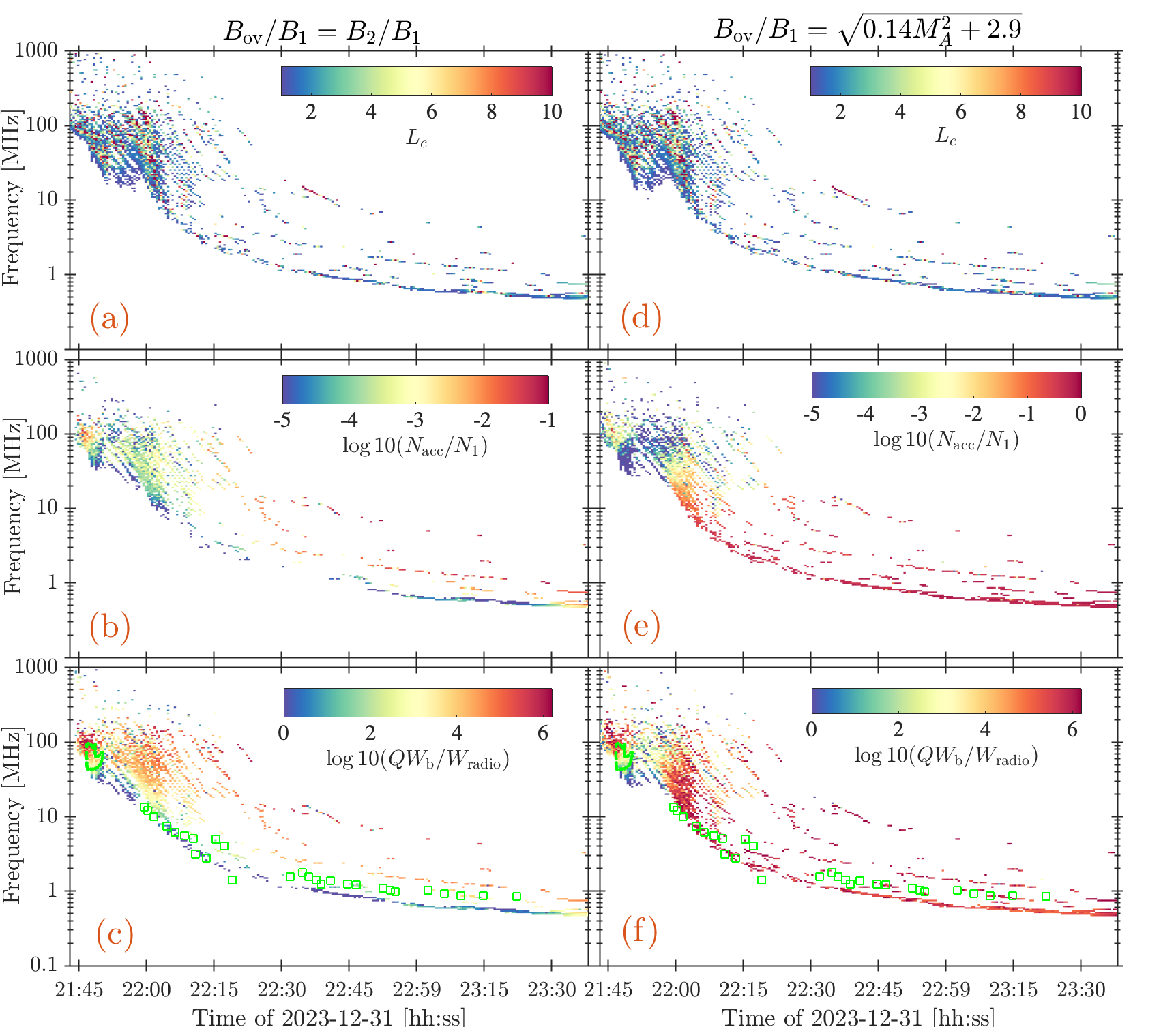}}
\small
  \caption{Same as \autoref{fig:synthetic_spectrogram}(panel a), here presenting the instability criterion $L_c$ (panels a and d), accelerated electron fraction $N_{\rm acc} / N_1$ (panels b and e), and efficiency of conversion of beam kinetic energy into radio waves $\varepsilon_{\rm rad}$ (panels c and f). Regions where instability is not triggered, i.e. $L_c<1$, were excluded in all panels. The left side panels \emph{a-c} employed the loss-cone angle $\alpha_{\rm lc}$ value with magnetic overshoot ratio $B_{\rm ov} / B_1 = B_2/B_1 = X_M$  and the right-side panels used: $B_{\rm ov} / B_1 =  \sqrt{0.14 M_A^2 + 2.9}$. In the bottom panels \emph{c} and \emph{f} the locations of the ``hotspot'' and DH type II burst signatures observed by Wind/WAVES (from analysis of \autoref{fig:Radio_PSP_STA_Wind} and \autoref{fig:Radio_eCallisto}) using green dotted line and green squares, respectively.
  }
\label{fig:SDA_spectra}
\end{figure}

These three quantities were estimated over the reconstructed evolution of the shock surface and casted into the synthetic spectrum, following the same method as for $M_A$ in \autoref{fig:synthetic_spectrogram} (panel a). We present the result of this analysis in \autoref{fig:SDA_spectra}. The $L_c$, $N_{\rm acc}/N_1$, and $\varepsilon_{\rm rad}$ parameters are presented in separate panels from top to bottom, respectively. The left side panels \emph{a-c} employed the loss-cone angle $\alpha_{\rm lc}$ with magnetic ratio $B_{\rm ov} / B_1 = B_2/B_1 = X_M$ as in \citet{Mann_2018,Mann_2022} studies and the right-side panels used the theoretical expression of \citet{Gedalin_2021} further applied and fitted to in-situ data of shock crossings by \citet{Lindberg_2025} : $B_{\rm ov} / B_1 = \sqrt{0.14 M_A^2 + 2.9}$. In the bottom panels \emph{c} and \emph{f} we also over-plotted the locations of the hotspot and DH type II burst observed by Wind/WAVES (from analysis of Figures 2 and 3) using greed dotted line and green squares, respectively.

Regions with $L_c<1$ were excluded in all panels of \autoref{fig:SDA_spectra}, in order to see only the instability-triggering parts of the shock. As can be seen in all panels, only few regions of the shock surface are able to produce fast enough electron beam. Late in the evolution, after 22:20 UT, these regions are particularly rare as the shock becomes mostly quasi-parallel except inside the interaction parts with the HCS \footnote{We also explicitly tested the importance of quasi-perpendicular shocks by excluding all $\theta_{\rm Bn} <70^\circ$ regions. It resulted in total disappearance of $L_c >1$ regions in all cases. Thus confirming that only the shock parts with $\theta_{\rm Bn} >70^\circ$ are sourcing sufficiently fast beams that lead to radio emission.}. By comparing two top panels \emph{a} and \emph{d}, we remark that $L_c$ value appears to be weakly sensitive to the choice of $B_{\rm ov} / B_1$ (i.e. value of $ \alpha_{\rm lc}$). It is expected as $\cos \alpha_{\rm lc}$ factor changes only slightly the value of  $L_c$.

The picture is very different for $N_{\rm acc} / N_1$ (panel b vs panel e). For the choice of $B_{\rm ov} / B_1 = B_2 / B_1$ (panel b) as typically done in Mann et al works, the values of $N_{\rm acc} / N_1$ evolve in range between $10^{-6}$ and $0.1$. While for $B_{\rm ov} / B_1 = \sqrt{0.14 M_A^2 + 2.9}$, the values are similar to the former before 21:55~UT but become nearly equal to $1$ after $\sim$22:00~UT. Thus, implying that all incoming electrons are reflected at the shock front which is highly questionable physically. 

Finally, the radio emissivity parameter $\varepsilon_{\rm rad}$ is presented in panels c and f. It is expected to directly compare with observations. The emissivity is very high early on in both cases, but drops significantly after 21:50~UT in panel c, while remaining very high in panel f. Some tiny regions of the shock surface appear to be able to generate $\varepsilon_{\rm rad} > 10^4$ inside the intermediate and upper shock-HCS interaction regions, while the lower main shock-HCS line has $\varepsilon_{\rm rad} \sim 1  - 10$. 
In the case of $B_{\rm ov} / B_1 = \sqrt{0.14 M_A^2 + 2.9}$ (panel f) all regions have $\varepsilon_{\rm rad} > 10^5$ since they mostly correspond to high-$M_A$ and $N_{\rm acc} / N_1 \simeq 1$ regions.

For visual comparison, we over plotted in bottom panels the ``hotspot'' region from \autoref{fig:Radio_eCallisto}(panel \emph{a}) using green dotted line and DH type II signatures using green square symbols, extracted from Wind/WAVES observations in \autoref{fig:Radio_PSP_STA_Wind}. Early evolution compares well with $\varepsilon_{\rm rad}$ distribution. The ``hotspot'' region between 21:44~UT and 21:47~UT is nearly in overlap with observed region, in green. Yet, the most intense modeled emission occurs slightly earlier and at slightly higher frequencies than observed. Between 22:00~UT and 22:15~UT the overlap between synthetic and observed emission regions is particularly good, i.e. when shock-HCS interaction emerges (see panels \emph{e} and \emph{f} in \autoref{fig:synthetic_spectrogram}). After 22:30~UT it appears that the intermediate shock-HCS interaction line matches better observed type II signatures despite being more intermittent but emitting at higher values of $\varepsilon_{\rm rad}$ than the main shock-HCS line that evolves at slightly lower frequencies with regular pattern.

\section{Discussion and conclusion}
\label{Sect:Conclusion} 

The results of our work inferred efficient acceleration of energetic electrons by the transient high-Mach quasi-perpendicular shock in the low corona. 
By examining EUV observations we focused on a particular feature ahead of the shock wave between 21:47 UT and 21:53 UT (see \autoref{fig:EUV_C2_obs} upper row), resembling an illuminated flux tube. By examining radio observations there was radio emission between 21:46 UT and 21:49 UT in 40 MHz to 140 MHz frequency range with reverse drifting features (as shown in \autoref{fig:Radio_eCallisto}). By reconstructing the shock front evolution (\autoref{Sect:3D_Reconstruction}) we then found in \autoref{Sect:Results} that the EUV flux tube feature and radio emission can both be attributed to the transient high-$M_A$ quasi-perpendicular shock patch on the Earth-sided flank of the coronal pressure wave. We then demonstrated in \autoref{subsect:SDA_synthetic_spectra} that this region was highly efficient at accelerating electrons by SDA with enough energy density to power radio emission. The same analysis presented \autoref{fig:SDA_spectra} of late evolution indicated good agreement between regions able to drive electron beam instability at shock, i.e. $L_c > 1$, and drifting DH type II burst.

There are, however, some discrepancies between our modeling and observations. Firstly, reconstructed field lines that connect the hotspot to the solar surface are similar but do not follow exactly the apparent topology of EUV flux tube. Modeled lines are few degrees closer to the solar equator. These lines are open filed lines in agreement with 193\angstrom EUV maps where shaded regions were seen to be illuminated. An important test of electron acceleration efficiency by this shock region would have been possible if direct magnetic connection of a spacecraft to this region occurred \citep[as done in, e.g.,][for a different event]{Morosan_2024}. Unfortunately, it did not happen as the spacecraft fleet was not positioned favorably. All available probes were connected far away.  Secondly, there was about one minute difference between modeled and observed radio emission by MEXART, as seen in \autoref{fig:Radio_eCallisto}(a). 

Despite these differences, it is remarkable that the plasma emission from the shock-hotspot region falls into consistent range of frequencies and traced field lines visually resemble observed flux tube in EUV during the correct time range. In some reconstruction results we found that the hotspot was sliding from higher to lower heights with time, potentially explaining the reverse drifting radio emission of one minute duration between 21:48 UT and 21:49 UT. The difference in onset time and field line topology discrepancy could be due to two main reasons: (i) shock fitting did not capture properly the shock position at relevant time and (ii) MAST simulation did not capture the position and orientation of the pseudo-streamer with required accuracy. For the former, we tried several strategies of shock fittings. In some cases largely exaggerating the expansion speed and main direction of propagation. None on these tries was successful in reproducing both the onset time and frequency range of hotspot-related emissions. We hence favor the latter explanation since even a slight shift (few degrees) of pseudo-streamer position could account for one minute delay between modeling and observation. We also expect that a slight southward shift of the pseudo-streamer’s cusp would improve agreement between hotspot-connected field lines and EUV-brightened flux tube.

We did not apply any correction factor on magnetic field intensity and plasma density from MAST MHD simulations. Indeed, it was common in previous studies to apply some correction factor (constant in the whole box), based on comparison between open magnetic flux in the simulation and field intensity measured in-situ \citep[][]{Rouillard_2016, Kouloumvakos_2019, Kouloumvakos_2021}. For densities, correction factors between 0.5 and 3 were applied by \citet{Kouloumvakos_2019, Kouloumvakos_2021, Morosan_2024}. Despite observational motivations \citep{Wang_2017}, we did not follow this procedure because (1) applying a constant correction factor everywhere in the simulation box will only improve some localized regions and (2) rescaling by hand one physical quantity modifies in principle the pressure balance in non-trivial way. In the present study, we noticed that rescaling densities by a factor of 2 will make the main shock-HCS interaction region correspond nearly perfectly to the observed type II burst pattern by STA and Wind (up-shift of lower dotted white line in \autoref{fig:Radio_PSP_STA_Wind}b,c). But at the same time it will shift the ``hotspot''-related emission to too high values as compared to observations. We conclude that methodological improvements are necessary for both shock reconstruction and steady-state coronal MHD simulations if one wants to capture small scale features (but still much larger than kinetic).

In terms of shock physics, we studied which parts of evolving shock surface were efficient at producing energetic electrons and driving radio emissions in \autoref{subsect:SDA_synthetic_spectra}. Two different prescriptions for the value of magnetic overshoot at the shock front were adopted. The most simple assumption that the overshoot value is equal to downstream value ($B_{\rm ov} = B_2$) provided sufficient level of radio emissivity at times and frequencies consistent with those of the observed type II radio bursts. On the other hand, the prescription $B_{\rm ov} / B_1 =  \sqrt{a M_A^2 + b}$ \citep{Gedalin_2021} while being motivated theoretically and by in-situ measurements near 1 AU and Saturn's bow shock \citep{Jebaraj_2023b, Lindberg_2025}, leads to nonphysically large values of reflected electron densities at the shock when the Alfv\'en Mach number is particularly high ($M_A >10$)\footnote{We also tried all different values of $a$ and $b$ from \citep{Lindberg_2025} study, obtaining too large $N_{\rm acc} / N_1 \sim 1$ in all cases.}. The issue with this prescription at large $M_A$ was already noticed by \citet{Lindberg_2025}, the data indicating flattening of $B_{\rm ov} / B_1$ instead of rising as $\propto M_A$. Our analysis of a fast shock inside solar corona shows that this prescription has to be revised for $M_A \gg 10$ shocks, especially regarding the requirement to fix the upper bound of $B_{\rm ov} / B_1$ in this regime.

\subsection{Conclusion}

To conclude, in the present work we studied several aspects of the coronal shock from the 31st December 2023 eruption. We focused on observational reconstruction of the coronal shock evolution and its role in producing radio emissions. By field line tracing and producing a synthetic radio spectrum we found that the short-lived radio emissions at $\sim100$ MHz can be explained by a transient ``hotspot'' (high Mach number and quasi-perpendicular shock at the cusp of pseudo-streamer) region that also reproduces rarely observed upstream flux tube illumination seen in EUV images. Our methodology was also efficient in reproducing the DH type II burst evolution as shock-HCS interaction regions, consolidating similar conclusion of previous studies \citep[e.g.,][]{Cho_2008, Jebaraj_2021, Frassati_2022, Koval_2023}. We also note that very similar conclusions were found in recent study of \citet{Bhandari_2025} that used similar methodology to the present work for ten events of Solar Cycle 25. Our methodology merits more general application to other type II bursts and related coronal shock waves in order to assess the general relevance of shock--pseudo-streamer and shock--HCS interaction regions as important places of significant particle acceleration and dominant emitting regions of type II bursts.

%

%

%
\begin{acks}
We are grateful to the anonymous referee for detailed analysis and insightful suggestions that substantially improved the manuscript. I.P. is grateful to K.L. Klein for insightful discussions and suggestions and Marco Pinto for providing Bepi Colombo/BERM data (not presented in the main text).
I.P. acknowledges the European Union's Horizon-2020 research and innovation program for support under grant agreement No. 101004159 (SERPENTINE). This article reflects only the authors’ view and the European Commission is not responsible for any use that may be made of the information it contains.
A.K. acknowledges financial support from NASA’s LWS grant 80NSSC25K0130 and HGIO grant 80NSSC24K0555.
We acknowledge the e-CALLISTO; Wind/WAVES; STEREO; SOHO/LASCO; SDO/AIA; PSP teams and Predictive Science Inc. for providing open data access used in this study. The STEREO/SECCHI data are produced by a consortium of RAL (UK), NRL (USA), LMSAL (USA), GSFC (USA),MPS (Germany), CSL (Belgium), IOTA(France),and IAS(France). SOHO is a mission of international cooperation between ESA and NASA.The SDO/AIA data used are courtesy of SDO (NASA) and the AIA consortium. 
We acknowledge usage of the tools made available by the French plasma physics data center (Centre de Données de la Physique des Plasmas; CDPP; \url{http://cdpp.eu/}), CNES and the space weather team of Toulouse at Solar-Terrestrial Observations and Modellings Service (STORMS; \url{https://stormsweb.irap.omp.eu/}). In particular, present study used the Propagation Tool (\url{http://propagationtool.cdpp.eu}), and the Magnetic Connectivity Tool (\url{http://connect-tool.irap.omp.eu/}).
This research has also made use of PyThea v1.0.0, an open-source and free Python package to reconstruct the 3D structure of CMEs and shockwaves \citep{Kouloumvakos_2022b}. The LASCO/CME catalog is generated and maintained at the CDAW Data Center by NASA and The Catholic University of America in cooperation with the Naval Research Laboratory.
The work at AIP carried out by A.P.R. was supported by the Alexander von Humboldt foundation.
I.C.J. acknowledges support from the Research Council of Finland (X-Scale, grant No.~371569). 
\end{acks}
%
%
%
%

\begin{dataavailability}
SDO/AIA data is publicly available at \url{https://aia.lmsal.com/}.
CME and pace-based radio burst data is publicly available at the CDAW Data Center (\url{https://cdaw/gsfc.nasa.gov}). The ground-based radio data are available at \url{https://e- callisto.org}.
Predictive Science Inc. simulations are publicly available at \url{https://www.predsci.com/mhdweb/data_access.php}.
Coronal shock reconstructions (table of fits) is available on request to the corresponding author.
\end{dataavailability}
%
%

%
%
 \bibliographystyle{spr-mp-sola}
 \bibliography{biblio}  

\end{document}